\documentclass[pra,twocolumn,amsmath,amssymb,floatfix]{revtex4}
\usepackage{amssymb}
\usepackage{amsmath}
\usepackage{graphicx}
\usepackage{epsfig}
\usepackage{hyperref}
\hypersetup{colorlinks=true}
\usepackage{braket}

\usepackage[small,bf]{caption}
\usepackage{verbatim}
\usepackage{subfig}
\usepackage{float}
\graphicspath{{./Figures/}}

\usepackage{bm}
\newcommand{\be}{\begin{equation}}
\newcommand{\ee}{\end{equation}}

\newcommand{\beq}{\begin{eqnarray}}
\newcommand{\eeq}{\end{eqnarray}}

\def\H1{\widehat{H}_1}

\DeclareMathOperator{\Tr}{Tr}

\begin{document}

\title[]{Integrability and duality in spin chains}

\author{Eyzo Stouten$^{1,2}$, Pieter W. Claeys$^{1,3,4}$, Jean-S\'{e}bastien Caux$^{1}$ and Vladimir Gritsev$^{1,5}$ }

\address{$^1$Institute for Theoretical Physics, Universiteit van Amsterdam, Science Park 904,
Postbus 94485, 1098 XH Amsterdam, The Netherlands\\
$^2$Fachbereich Physik, Bergische Universit\"{a}t Wuppertal, 42097 Wuppertal, Germany\\
$^3$Department of Physics and Astronomy, Ghent University, Krijgslaan 281-S9, 9000 Ghent, Belgium\\
$^4$Center for Molecular Modeling, Ghent University, Technologiepark 903, 9052 Ghent, Belgium \\
$^5$LPTMS, CNRS, Univ. Paris-Sud, Universit\'{e} Paris-Saclay, 91405 Orsay, France}
\begin{abstract}
We construct a two-parametric family of integrable models and reveal their underlying duality symmetry.
A modular subgroup of this duality is shown to connect non-interacting modes of different systems. We apply this solution and duality to a Richardson-Gaudin model and generate a novel integrable system termed the $s$-$d$ wave Richardson-Gaudin-Kitaev interacting chain, interpolating  $s$- and $d$- wave superconductivity. The phase diagram of this model has a topological phase transition that can be connected to the duality, where the occupancy of the non-interacting mode serves as a topological order parameter.
\end{abstract}


\maketitle


{\it Introduction.} Integrable models (IM) play a crucial role in our understanding of low-dimensional statistical systems and condensed matter physics. One fundamental example of exactly-solvable many-body systems is the model of a one-dimensional Bose gas interacting via a short-range (delta-function) contact potential, the so-called Lieb-Liniger (LL) model \cite{LL1,LL2}. Several remarkable properties of this model were subsequently obtained since its discovery in the sixties: the absence of Bose-condensation for any interaction strength, effective fermionization at infinitely large repulsion strengths, the power-law correlations at low energies as indication of the Luttinger liquid behavior and the absence of thermalization in non-equilibrium setups \cite{Cazalilla-rev}. Various predictions of the Bethe ansatz solution of this model have also been extensively checked experimentally in cold atomic systems \cite{KWW1,KWW2,Bloch,Ams,DSF,Meinert}, and applications of this integrable model currently go far beyond the physics of 1D cold gases \cite{Lai1,Lai2,Chang,Imamoglu,Kardar,CD,Dotsenko} 

However, the LL model does not exhaust all integrable continuum models in 1D. In particular, the Yang-Gaudin model generalizes it to interacting fermions \cite{gaudin_bethe_2014}, and multicomponent mixtures of point-interacting models have been obtained by Sutherland \cite{S}. In a set of remarkable papers, C. N. Yang also generalized the LL model to particles in arbitrary representation of the symmetric group \cite{Yang1,Yang2}. These models allow for integrable generalizations when an isolated impurity is coupled to interacting fermions. This gave rise to solvable Kondo and Anderson models \cite{Kondo} which were instrumental in understanding the correctness of the numerical renormalisation group of Wilson and Kogut \cite{WK}, and allowed to obtain exact expressions for observables and paved the way for the physics of non-Fermi liquid systems \cite{non-FL}.

The models mentioned above are defined in a continuum, while important models on a one-dimensional lattice play, perhaps, an even more fundamental role. In particular, the Hubbard model \cite{Hubbard} and its large-interaction limit, the spin-1/2 Heisenderg model \cite{gaudin_bethe_2014}, were instrumental in understanding the nature of quantum phase transitions in low-D, fractionalized excitations, spin-charge separation and the importance of topological phenomena. Eventually, accumulation of this knowledge resulted in the {\it universal} paradigm of Luttinger liquids \cite{Haldane}, a low-D counterpart of the concept of a Fermi liquid.

Another broad class of solvable many-body models, the so-called Richardson-Gaudin models \cite{richardson_restricted_1963,richardson_exact_1964,gaudin_bethe_2014,dukelsky_colloquium:_2004,ortiz_exactly-solvable_2005},  can be obtained as a particular limit of spin-1/2 model where inhomogeneities are allowed. This class of models embraces the Tavis-Cummings and Dicke models of superradiance \cite{TCD}, Richardson's reduced BCS model of superconductivity \cite{richardson_new_2002,richardson_restricted_1963,richardson_exact_1964,amico_integrable_2001}, central spin model of electrons in quantum dots and nitrogen vacancy centers in diamond \cite{QD}, Lipkin-Meshkov-Glick models of nuclei \cite{lerma_h._lipkinmeshkovglick_2013}, and many more. This class of models is popular because of their relevance for solid-state based quantum computation, quantum decoherence, quantum information and excitation energy transfer.

Integrable models have attracted increasing interest in contemporary field and string theory. This interest emerged in a 2D context \cite{ZZ}, and independently in 3D with invariants of knots \cite{W-knots}, continuing in 4D with the AdS/CFT correspondence \cite{Beisert} and supersymmetric gauge theories \cite{NS}, and led to contemporary developments in Ref. \cite{W1}. Dualities in various forms  were always a close companion of these developments \cite{string-dual}. The recent echo of these ideas into the realm of condensed matter physics \cite{Seiberg} may become a powerful tool for better understanding quantum criticality, correlated and topological states of matter \cite{Feiguin}.

Motivated by these developments, we here investigate a basic integrable model from the duality point of view and reveal a symmetry which is reminiscent of the duality in certain string theories \cite{Witten2} and the fractional Hall effect \cite{Dolan}. Using this duality we reveal new integrable models and, studying a particular case in more detail, find a topological phase and phase transition.

{\it Integrability and the Yang-Baxter equation.}
A key ingredient of the integrability of all these models is the fact that any three-body scattering factorizes into two-body scatterings, which obey the Yang-Baxter equation (YBE). Individual two body scatterings are parameterized by $S$- or $R$-matrices where the latter used in the case of integrable lattice models. Integrability conditions on the $R$-matrices are given by the celebrated Yang-Baxter equation,
$R_{12}(u,v)R_{13}(u,w)R_{23}(v,w)= R_{23}(v,w) R_{13}(u,w) R_{12}(u,v)$,
which expresses a factorization property of scattering between three particles (labeled by $1,2,3$) with respective rapidities $u,v,w$. The simplest known solution, suggested by Yang \cite{Yang1,Yang2}, is the {\it rational} $R$-matrix given by
\beq
R_{12}(u,v) = b(u,v)I_{12} + c(u,v) P_{12},
\eeq
where $I_{12}$ and $P_{12}$ are the identity and permutation operators acting in the direct product of the two Hilbert spaces for particles $1$ and $2$, with the {\it rational} functions defined as
\begin{equation}
b(u,v) = \frac{u-v}{u-v+\eta}, \qquad c(u,v) = \frac{\eta}{u-v+\eta},
\label{old}
\end{equation}
depending on a free parameter $\eta \in \mathbb{C}$ and satisfying  $b(u,v)+c(u,v)=1$. The Richardson-Gaudin limit is defined as $\eta\rightarrow 0$, where the inclusion of inhomogeneities can be used to obtain a non-trivial limit.

In this paper we are interested in the {\it general two parameter solution} for the $R$-matrix of the rational form. For that we parametrize it as \footnote{This is possible because the solution of the Yang-Baxter equation is defined up to multiplication by an arbitrary function of rapidities.}
\beq
R_{12}(u,v)=I_{12}+F(u,v)P_{12}.
\label{R-m}
\eeq
Substituting this ansatz into the YBE leads to the following functional equation for $F(u,v)$
\beq
F(u,v)F(u,w)+F(u,w)F(v,w)=F(u,v)F(v,w).
\label{F-eq}
\eeq
Manifestly, the well-known form (\ref{old}) for $F(u,v)\equiv c(u,v)/b(u,v)=\eta/(u-v)$ is a direct solution of this equation. We found a {\it general two parameter solution} \footnote{It was brought to our attention by the author of \cite{Martins} that a more general rational solution to the Yang-Baxter equation exists and that the solution which was independently derived in this publication is in fact a special case of the solution described in Ref. \cite{Martins}.} to Eq.~(\ref{F-eq}) in the class of {\it rational } functions. This solution is parametrized by two free couplings and could generate new integrable many-body 1D systems with possibly as many physical applications as the Lieb-Liniger or spin models before. The same functional equation as in Eq. (\ref{F-eq}) defines a class of Richardson-Gaudin models, which can be used to illustrate the possibilities of this parametrization.

{\it Generalization of the known solution.}
Since we are looking for a rational generalization of Eq. (\ref{old}), the following {\it ansatz} is considered for the function $F(u,v)$
\beq
F(u,v)=\frac{1}{u-v}\sum_{p,q=0}^{N}c_{p,q} u^{p}v^{q},
\label{cpq}
\eeq
where the coefficients $c_{p,q}$ are to be determined. Here the order $N$ of the polynomial is assumed to be finite \footnote{Generalizations for $N=\infty$ would lead to generalizations of the trigonometric/hyperbolic/elliptic classes and are postponed to the future.}.

The {\it first} observation is the following. A combination of analytical and numerical checks leads to the following general solution of Eq.~(\ref{F-eq})
\beq
F(u,v)=\frac{c_{0}^{2}+c_{0}c_{1}(u+v)+c_{1}^{2}uv}{ u-v},
\label{F-new}
\eeq
where $c_{0}$ and $c_{1}$ are {\it free} parameters (trivially related to $c_{p,q}$ in (\ref{cpq})). Polynomials of higher order in $u,v$ do not provide a solution to Eq. (\ref{F-eq}).

{\it $SL(2)$ duality of integrable models. }
The {\it second} important observation is an intrinsic $SL(2)$ duality symmetry associated with this solution. Namely, one can notice that if the rapidities $u$ and $v$ of the matrix (\ref{R-m}) are transformed according to the fractional-linear conformal transformation $(k\equiv (u,v))$,
\beq
\tilde{k}=\frac{\alpha k +\beta}{\gamma k+\delta},\qquad \alpha\delta-\gamma\beta=1,
\label{k-symm}
\eeq
the solution (\ref{F-new}) remains the same {\it iff} the couplings $c_{0}$ and $c_{1}$ are simultaneously transformed as
\beq
\left(
\begin{array}{c}
  \tilde{c}_{0}   \\
 \tilde{c}_{1}
\end{array}
\right)=\left(
\begin{array}{cc}
  \delta   & \beta\\
 \gamma & \alpha
\end{array}
\right)\left(
\begin{array}{c}
  c_{0}   \\
 c_{1}
\end{array}
\right).
\label{c-symm}
\eeq
Here, the unimodularity condition $ \alpha\delta-\gamma\beta=1$ is essential. The rapidities $k=(u,v)$ are, in principle, allowed to take arbitrary complex values, so the parameters $\alpha,\beta, \gamma,\delta$ could be complex as well, thus transforming under the group $SL_{k}(2,\mathbb{C})$ (the subscript $k$ denotes that they act on the rapidities). While at the moment the couplings $c_{0,1}$ can be considered complex as well, we however restrict ourselves to the real domain.

While this symmetry is preserved at the level of the scattering matrix and the YBE equation, it acts however nontrivially on a system with fixed external boundary conditions (say, periodic). The condition of single-valuedness of the wave function then leads to the Bethe equations, which are derived in the general form in the Supplementary material. The solutions of the Bethe equations are then not symmetric with respect to the transformations (\ref{k-symm}) and (\ref{c-symm}), but rather generate a {\it duality} between different models, since they relate different physical models. In this case, the $SL(2,\mathbb{R})$ symmetry is similar to the duality in string theory (e.g., in IIB superstring theory) relating theories with different coupling constants \cite{Witten2}.


It is important to note that there are special points in the space of rapidities when the scattering matrix trivializes and $R(u,v)\equiv 1$. This happens when some of the radidities satisfy $u= -c_{0}/c_{1}$. In this case the Bethe equations reduce to quantization conditions for non-interacting particles.
This observation combined with the $SL(2,\mathbb{R})$ duality implies that a subgroup of the latter, the {\it modular} group $SL(2, \mathbb{Z})$, connects, in a sense of duality transformation all free, noninteracting modes of the model.



{\it Richardson-Gaudin limit.}
Starting from a solution to the Yang-Baxter equation, exactly solvable models can be constructed containing either bosonic or spin degrees of freedom, depending on the representation of the algebra. In this regard, a solution to the functional equation (\ref{F-eq}) is also known to determine integrable Richardson-Gaudin spin systems \cite{richardson_restricted_1963,richardson_exact_1964,amico_integrable_2001,dukelsky_colloquium:_2004,ortiz_exactly-solvable_2005,gaudin_bethe_2014}. A class of integrable spin systems can be obtained by plugging in the solution (\ref{F-new}), 
and the influence of the $SL(2,\mathbb{R})$ symmetry can be made apparent.

Richardson-Gaudin models are characterized by a set of conserved charges
\beq\label{RG:com}
Q_i = S_i^z -  \sum_{j \neq i} F_{ij} \left[\frac{1}{2}\left(S_i^+S_j^-+S_i^-S_j^+\right)+ S_i^z S_j^z\right],
\eeq
with the demand $[Q_i,Q_j]=0$ leading exactly to Eq. (\ref{F-eq}) with $F_{ij}\equiv F(\epsilon_{i},\epsilon_{j})$ with arbitrary inhomogeneities $\epsilon_{i,j} \in \mathbb{R}$. Inserting Yang's solution then leads to the rational Richardson model, best-known in the context of superconductivity and the BEC-BCS crossover \cite{dukelsky_colloquium:_2004}, while the  solution (\ref{F-new}) leads to the rational limit of a parametrization proposed in [\onlinecite{richardson_new_2002}]. For this parametrization, the conserved charges are explicitly invariant under the $SL(2,\mathbb{R})$ symmetry. An integrable Hamiltonian can then be constructed by taking a linear combination of these conserved charges, satisfying $[H,Q_i]=0$ by construction. However, the choice of linear combination can break the $SL(2,\mathbb{R})$ invariance, similar to the choice of boundary conditions in the generic Bethe ansatz model, possibly leading to a non-trivial phase diagram. A large variety of resulting integrable Hamiltonians can be obtained depending on the parametrization and the choice of a Hamiltonian, but we will illustrate some of the resulting physics with a specific model based on the Richardson-Gaudin-Kitaev chain describing topological superconductivity in fermion chains \cite{ortiz_many-body_2014}.

By mapping the spin $su(2)$ algebras to fermion pairing operators, an integrable Hamiltonian can be found as
\beq
H = \sum_{k} \epsilon_kc^{\dagger}_k c_k - \frac{G}{2} (C^{\dagger}C+CC^{\dagger}) - \frac{G}{4} BB, \\
C^{\dagger} = \sum_{k>0} \eta_k c^{\dagger}_k c^{\dagger}_{-k}, \qquad B = \sum_{k} \eta_k c^{\dagger}_k c_k,
\eeq
for one-dimensional fermions with momentum $k$ and single-particle spectrum $\epsilon_k = -2t_1 \cos k - 2 t_2 \cos 2k $, and interactions modulated by $\eta_k= 4 \sin^2(k/2)(t_1 + 4 t_2\cos^2(k/2))$  with interaction strength $G$. These are related to the previous parametrization through $\sqrt{G} \eta_k = c_0+c_1 \epsilon_k$ with $2 (t_1+t_2) = c_0/c_1$ and $G^{-1} = 2 c_1^{-2} - \sum_k \eta_k$, see \cite{SM}.

\begin{center}
\begin{figure}
\includegraphics[width=\columnwidth]{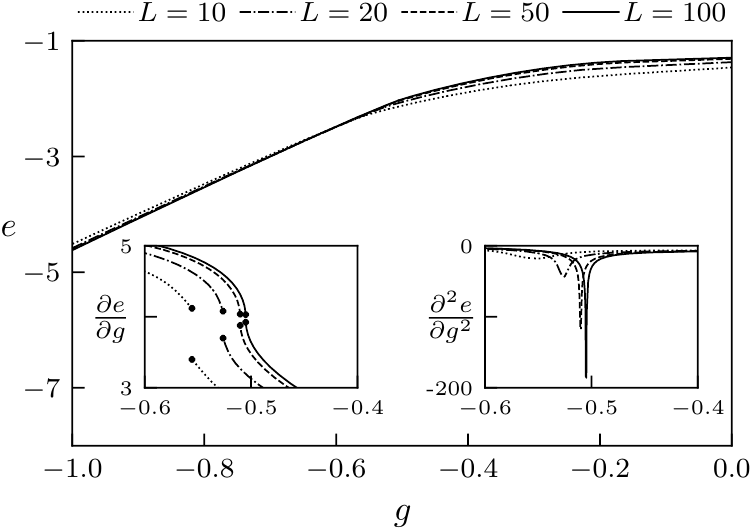}
\caption{The energy per site $e=E/L$ and its first and second derivative as a function of the interaction strength $g=GL$ for systems of length $L$ at half-filling and periodic boundary conditions. The system parameters are $t_1=1$ and $t_2=0$, with $g_c=G_c L$ marked by vertical lines for different system sizes and moving toward $g_c^{-1} = -2(t_1+t_2)$ in the thermodynamic limit (here $-1/2$). \label{fig:RG:ener}}
\end{figure}
\end{center}

This model is integrable for any choice of the momentum $k$-distribution, and describes a chain with either nearest neighbour-interactions ($t_2 =0$) or long-range interactions ($t_2 \neq 0$) in real space, on top of which pairing interactions ($C^{\dagger}C$) and long-range interactions ($B$) have been added. This model can be seen as a variant on the Richardson-Gaudin-Kitaev chain \cite{ortiz_many-body_2014} with pairing interactions interpolating between $s$- and $d$-wave pairing \cite{marquette_integrability_2013}, also motivating the choice of parametrization.
In the determination of the phase diagram, a crucial element is that the interaction vanishes at certain {\it finite} values of the momentum (here $k=0$, for which $\eta_k = 0$ and $\epsilon_k = -c_0/c_1$ by construction). Equivalently, this corresponds to trivial particle scattering with $F_{ij}=0$. 
The existence of non-interacting modes in Richardson-Gaudin models has previously been linked to the existence of Majorana fermions in particle-number conserving models \cite{ortiz_many-body_2014,ortiz_what_2016} and topological phase transitions \cite{ibanez_exactly_2009,rombouts_quantum_2010,marquette_integrability_2013}. This can be illustrated on the model at hand, where the ground state energy can be exactly obtained for large system sizes because of the exact solvability. In Fig.~(\ref{fig:RG:ener}) we plot the energy per site and its first and second derivative for varying interaction strength. For finite system sizes, the discontinuity in the first derivative points to a first-order quantum phase transition. In the thermodynamic limit this discontinuity vanishes and the second derivative diverges, indicating a second-order phase transition, consistent with recent results on the thermodynamics of topological phases \cite{kempkes_universalities_2016}.
Furthermore, this is a phase transition between a topologically non-trivial phase ($G>G_c$) and a topologically trivial phase ($G<G_c$). This can be obtained from mean-field theory in the thermodynamic limit [\onlinecite{marquette_integrability_2013}], or from the recently-proposed characterization of topological superconductivity in number-conserving systems  for finite system sizes \cite{ortiz_many-body_2014,ortiz_what_2016}. Following this last route, the occupation of the single-particle levels for vanishing momentum $k$ can be seen as the one-dimensional equivalent of a {\it winding number} \cite{foster_quantum_2013,ortiz_many-body_2014,ortiz_what_2016}, and it can be checked that $\lim_{k \to 0 }\hat{n}_{k} = 1$ in the topological phase and  $\lim_{k \to 0 }\hat{n}_{k} = 0$ in the trivial phase.

This transition can be understood through symmetry arguments. The Hamiltonian undergoes a quantum phase transition at $G_c^{-1}=-\sum_k \eta_k$, which can be mapped back to a limit where $c_1 \to \infty$ while $c_0/c_1$ remains fixed. In this limit, the conserved charges reduce to those of the Gaudin model, which has an additional $su(2)$ total spin-symmetry compared to the Richardson-Gaudin models \cite{gaudin_bethe_2014}. This symmetry results in level crossings in the spectrum between states with different particle numbers (spin-projection). Combined with the existence of a non-interacting level, this leads to level crossings between states with the same particle number at the symmetric point, since changing the occupation of the non-interacting level does not influence the energy. The quantum phase transition is then caused by a level crossing of two levels with different occupations of this level, leading to a vanishing chemical potential and a change in topological invariant \cite{foster_quantum_2013}. Similarly, the rapidity distribution in complex space also undergoes a transition, as shown in the supplementary material and which is expected for a quantum phase transition \cite{links_ground-state_2015}. The $SL(2,\mathbb{R})$ symmetry can again be compared to the field theory duality which relates theories with different coupling constants, since the ground state can be mapped to eigenstates of the Richardson model in different regimes ($c_1^2$ and $c_0^2$ change sign at the phase transition). It should be stressed that the existence of such a transition does not depend on the specific choice of Hamiltonian, but is a fundamental property of the new parametrization (with $c_1 \neq 0$) and the resulting combination of total spin $su(2)$-symmetry and the $SL(2,\mathbb{R})$ symmetry.

{\it Discussion and conclusions}.
In this paper we presented a general rational solution of the Yang-Baxter relation in the fundamental (spin-1/2) representation. A hidden duality of this solution was demonstrated, connecting different models, and a modular subgroup $SL(2,\mathbb{Z})$ of this duality connects special non-interacting modes of different systems. Constructing a Richardson-Gaudin model from this solution, it was shown how in these models the non-interacting modes serve as a topological order parameter and are responsible for a topological phase transition. While here we mostly focused on a Gaudin-Richardson limit of this solution it can be directly extended to inhomogeneous spin chains, Lieb-Liniger type models, and higher spin models.

{\it Acknowledgements.} The authors would like to thank M. Zvonarev and A. Polychronakos for their stimulating discussions. Also we are grateful to M. Martins for his correspondence. This work is part of the Delta-ITP consortium, a program of the Netherlands Organization for Scientific Research (NWO) that is funded by the Dutch Ministry of Education, Culture and Science (OCW). P.W.C. acknowledges support from a Ph.D. fellowship and a travel grant for a long stay abroad at the University of Amsterdam from the Research Foundation Flanders (FWO Vlaanderen).


\newpage
\clearpage
\onecolumngrid

\renewcommand{\theequation}{S\arabic{equation}}
\renewcommand{\thepage}{S\arabic{page}}
\renewcommand{\thesection}{S\arabic{section}}
\renewcommand{\thetable}{S\arabic{table}}
\renewcommand{\thefigure}{S\arabic{figure}}
\renewcommand{\bibnumfmt}[1]{[{\normalfont S#1}]}
\setcounter{page}{0}
\setcounter{equation}{0}

\begin{center}
{\bf \Large Supplementary Material }
\end{center}

\section{A most general rational solution for the Yang-Baxter Equation: Algebraic Bethe Ansatz and Richardson-Gaudin limit }
\subsection*{Algebraic Bethe Ansatz}
In order to be self-contained, we first derive the general form of Bethe equations for the model under study and show how to derive Richardson-Gaudin integrability starting from a solution to the Yang-Baxter equation \cite{korepin_quantum_1993}. Given such an $R$-matrix parametrized as
\begin{equation}
R_{12}(u,v) = \frac{I_{12}+\eta F(u,v) P_{12}}{1+\eta F(u,v)} =
\begin{pmatrix}
1 & 0 & 0 & 0 \\
0 & b(u,v) & c(u,v) & 0 \\
0 & c(u,v) & b(u,v) & 0 \\
0 & 0 & 0 &1
\end{pmatrix},
\end{equation}
with $b(u,v) = 1/(1+ \eta F(u,v))$ and $c(u,v) = \eta F(u,v)/(1+\eta (u,v))$, the building block of any integrable model is a Lax operator satisfying
\begin{equation}\label{app:RGint:RLL}
R_{12}(u,v) L_{j,1}(u)L_{j,2}(v) = L_{j,2}(v)L_{j,1}(u) R_{12}(u,v),
\end{equation}
from which a monodromy matrix $T_a(u)$ and a transfer matrix  $\tau(u)$ can be constructed as
\begin{equation}
T_a(u) = L_{L,a}(u)\dots L_{2,a}(u) L_{1,a}(u) = \begin{pmatrix}
A(u) & B(u) \\
C(u) & D(u)
\end{pmatrix}_a
, \qquad \tau(u) = \Tr_a \left[T_a(u)\right] = A(u)+D(u),
\end{equation}
where the indices $i=1 \dots L$ denote the physical Hilbert spaces, and $a$ is an auxiliary space which is traced over (this construction corresponds to periodic boundary conditions). This gives rise to a continuous set of commuting operators
\begin{equation}
[\tau(u), \tau(v)] = 0, \qquad \forall u,v \in \mathbb{C}.
\end{equation}
Following the usual techniques of the Algebraic Bethe Ansatz, the parametrization of the $R$-matrix in terms of $b(u,v)$ and $c(u,v)$ allows results to be directly transferred to the present situation \cite{korepin_quantum_1993}. The eigenstates of $\tau(u)$ are given by Bethe states
\begin{equation}
\ket{\{\lambda\}_N} = \prod_{j=1}^N B(\lambda_j)\ket{0},
\end{equation}
acting on a vacuum state satisfying $C(u)\ket{0}=0$, $A(u)\ket{0}=a(u)\ket{0}$ and $D(u)\ket{0} = d(u)\ket{0}$, leading to eigenvalues
\begin{equation}
\tau(u|\{\lambda\}_N)=a(u)\prod_{k=1}^N \frac{1}{b(\lambda_k,u)}+ du)\prod_{k=1}^N \frac{1}{b(u,\lambda_k)},
\end{equation}
provided the Bethe equations are satisfied
\begin{equation}
\frac{a(\lambda_l)}{d(\lambda_l)}\prod_{k \neq l}^N \frac{b(\lambda_l,\lambda_k)}{b(\lambda_k,\lambda_l)} = 1, \qquad \forall l=1 \dots N.
\end{equation}

\subsection*{Richardson-Gaudin integrability}

The limit $\eta \to 0$ is known as the quasi-classical limit, leading to the class of integrable Richardson-Gaudin models \cite{von_delft_algebraic_2002,links_algebraic_2003}. This can be seen as a linearization of the usual construction, where a series expansion for the $R$-matrix in $\eta$ results in
\begin{equation}
R_{12}(u,v) = I_{12}-\eta \mathcal{R}_{12}(u,v) + \mathcal{O}(\eta^2),
\end{equation}
defining the quasi-classical $\mathcal{R}$-matrix as
\begin{equation}
\mathcal{R}_{12}(u,v) = F(u,v)(I_{12}-P_{12})= -2 F(u,v) \left(\vec{S}_1 \cdot \vec{S}_2 - \frac{1}{4}\right).
\end{equation}
A Lax operator can be obtained from the $R$-matrix as $L_{j,a}(u)=g_a R_{aj}(u,\epsilon_j)$ with $\epsilon_j \in \mathbb{R}$ and $g_a = e^{-2\eta S^z_a /  L}$ acting solely on the auxiliary space, satisfying the RLL-relation (\ref{app:RGint:RLL}) by construction. This operator can similarly be expanded in $\eta$ as
\begin{equation}
L_{j,a}(u) = I_{aj}-2\eta\left[\frac{S^z_a}{L} - F(u, \epsilon_j)\left(\vec{S}_a \cdot \vec{S}_j - \frac{1}{4}\right)\right]+\mathcal{O}(\eta^2).
\end{equation}
Taking the quasi-classical limit, the only non-trivial contributions to the transfer matrix will be at values $u=\epsilon_j$, where
\begin{align}
\tau(\epsilon_j)=&\Tr_a \left[L_{L,a}(\epsilon_j) \dots L_{j+1,a}(\epsilon_j) g_a \mathbb{P}_{aj} L_{j-1,a}(\epsilon_j) \dots L_{1,a}(\epsilon_j)\right] \nonumber \\
=& L_{j-1,j}(\epsilon_j) \dots L_{1,j}(\epsilon_j)L_{L,j}(\epsilon_j) \dots L_{j+1,j}(\epsilon_j) g_j,
\end{align}
due to the cyclic invariance of the trace. Retaining all first-order terms in $\eta$ in the quasi-classical limit then results in
\begin{equation}
\tau(\epsilon_j) = 1 - {2\eta}\left[S_j^z -   \sum_{k \neq j}^L F(\epsilon_j, \epsilon_k)\left(\vec{S}_j \cdot \vec{S}_k -\frac{1}{4}\right)\right]+ \mathcal{O}(\eta^2).
\end{equation}
It follows that the operators
\begin{equation}
Q_j =S_j^z - \sum_{k \neq j}^L F(\epsilon_j, \epsilon_k)\left(\vec{S}_j \cdot \vec{S}_k -\frac{1}{4}\right), \qquad j=1 \dots L,
\end{equation}
construe a set of conserved operators $[Q_i,Q_j]=0, \forall i,j=1 \dots L$, which are exactly the conserved charges defining a class of Richardson-Gaudin models.

Taking the $\eta \to 0$ limit of the Bethe states and Bethe equations, these conserved charges can again be immediately diagonalized with Bethe states
\begin{equation}
| \{\lambda\}_N \rangle = \prod_{j=1}^N \mathcal{B}(\lambda_j)|0 \rangle, \qquad \mathcal{B}(\lambda) = \sum_{j=1}^L F(\epsilon_j, \lambda) S_j^-, \qquad |0\rangle =  | \uparrow \dots \uparrow \rangle,
\end{equation}
where the eigenvalues follow as
\begin{equation}
\qquad Q_j | {\{\lambda\}_N}\rangle = \frac{1}{2}\left(1 -  \sum_{k=1}^N F(\epsilon_j,\lambda_k) \right)|{\{\lambda\}_M}\rangle,
\end{equation}
provided the rapidities satisfy the Bethe equations
\begin{equation}
-1+\frac{1}{2}\sum_{j=1}^L F(\epsilon_j, \lambda_k) - \sum_{l \neq k}^N F(\lambda_l, \lambda_k)=0, \qquad k=1 \dots N.
\end{equation}

\section{Deriving the Richardson-Gaudin Hamiltonian}
\label{app:RG}
Starting from the conserved charges (where a constant has been subtracted)
\beq
Q_i = \left(S_i^z+\frac{1}{2}\right) -  \sum_{j \neq i}^L F_{ij} \left[\frac{1}{2}\left(S_i^+S_j^-+S_i^-S_j^+\right)+ S_i^z S_j^z\right],\nonumber\\
\eeq
with $F_{ij} = F(\epsilon_i,\epsilon_j)=(c_0+c_1 \epsilon_i) (c_0+c_1 \epsilon_j)/(\epsilon_i-\epsilon_j) = G \eta_i \eta_j / (\epsilon_i-\epsilon_j)$ (with $\sqrt{G}\eta_i = (c_0+c_1 \epsilon_i)$), taking $H = \sum_i \epsilon_i Q_i $ results in
\beq
H = \sum_{i=1}^L \left(S_i^z+\frac{1}{2}\right) - \frac{G}{2}\sum_{i, j \neq i}^L\eta_i \eta_j \left[\frac{1}{2}\left(S_i^+S_j^-+S_i^-S_j^+\right)+ S_i^z S_j^z\right].\nonumber\\
\eeq
The Hamiltonian can now be rewritten by associating a one-dimensional (positive) momentum $k$ with each level and mapping the $su(2)$-algebra to a quasispin algebra, introducing the fermion pair operators $S_k^z =\frac{1}{2}\left(c^{\dagger}_k c_k + c^{\dagger}_{-k}c_{-k}-1\right) \equiv \frac{1}{2}\left(n_k+n_{-k}-1\right)$, $S_k^{+} = c^{\dagger}_{k} c^{\dagger}_{-k} \equiv b^{\dagger}_k$ and $S_k^{-} = c_{-k}c_{k} \equiv b_k$. This leads to a Hamiltonian
\begin{align}
\sum_{k} \epsilon_k Q_k &=  \frac{1}{2}\sum_{k} \epsilon_k \left(n_k+n_{-k}\right) - \frac{G}{4} \sum_{k,k'}\eta_{k}\eta_{k'} \left(b^{\dagger}_{k}b_{k'}+b_{k}b^{\dagger}_{k'}\right) -\frac{G}{8}\sum_{k,k'} \eta_{k} \eta_{k'} \left(n_k+n_{-k}\right)\left(n_{k'}+n_{-k'}\right)  \nonumber\\
&\qquad \qquad \qquad \qquad +\frac{G}{8} \sum_{k,k'} \eta_{k}\eta_{k'}\left(n_k+n_{-k}+n_{k'}+n_{-k'}\right)+ Cst.
\end{align}
The sum over $k' \neq k$ has been extended to include $k=k'$ by adding and subtracting the Casimir operators of the spin algebras from the Hamiltonian, since these only lead to a global shift in the energy and can be absorbed in the constant term. By grouping terms together, we can get
\begin{align}
\sum_{k} \epsilon_k Q_k &= \frac{1}{2}\sum_{k} \epsilon_k \left(n_k+n_{-k}\right)\left[1+\frac{c_1}{2}\sum_{k'}(c_0+c_1 \epsilon_{k'})\right]  - \frac{G}{4} \sum_{k,k}\eta_{k}\eta_{k'} \left(b^{\dagger}_{k}b_{k'}+b_{k}b^{\dagger}_{k'}\right) \nonumber\\
&\qquad -\frac{G}{8}\sum_{k,k'} \eta_{k} \eta_{k'} \left(n_k+n_{-k}\right)\left(n_{k'}+n_{-k'}\right) +\frac{1}{4}\sum_{k'}\left( c_0+c_1 \epsilon_{k'}\right)\sum_{k} \left(n_k + n_{-k}\right) + Cst.
\end{align}
Since the total number of fermions $\sum_k (n_k+n_{-k})$ is a symmetry of the system, this term also reduces to a constant,  and this total operator can be divided by $\frac{1}{2}\left[1+\frac{c_1}{2} \sum_{k}(c_0+c_1 \epsilon_k)\right]$, resulting in the proposed Hamiltonian with $\epsilon_k = -2 t_1 \cos k - 2 t_2 \cos 2k$, $2 (t_1+t_2) = c_0/c_1$ and $G^{-1} = 2 c_1^{-2} - \sum_k \eta_k$. At the phase transition $c_1^{2} \to \infty$ with $c_0/c_1$ fixed, which is reflected in $G_c^{-1} = -\sum_k \eta_k$.

\section{Numerical solutions to the Richardson-Gaudin equations}
\begin{figure}
\begin{center}
\includegraphics{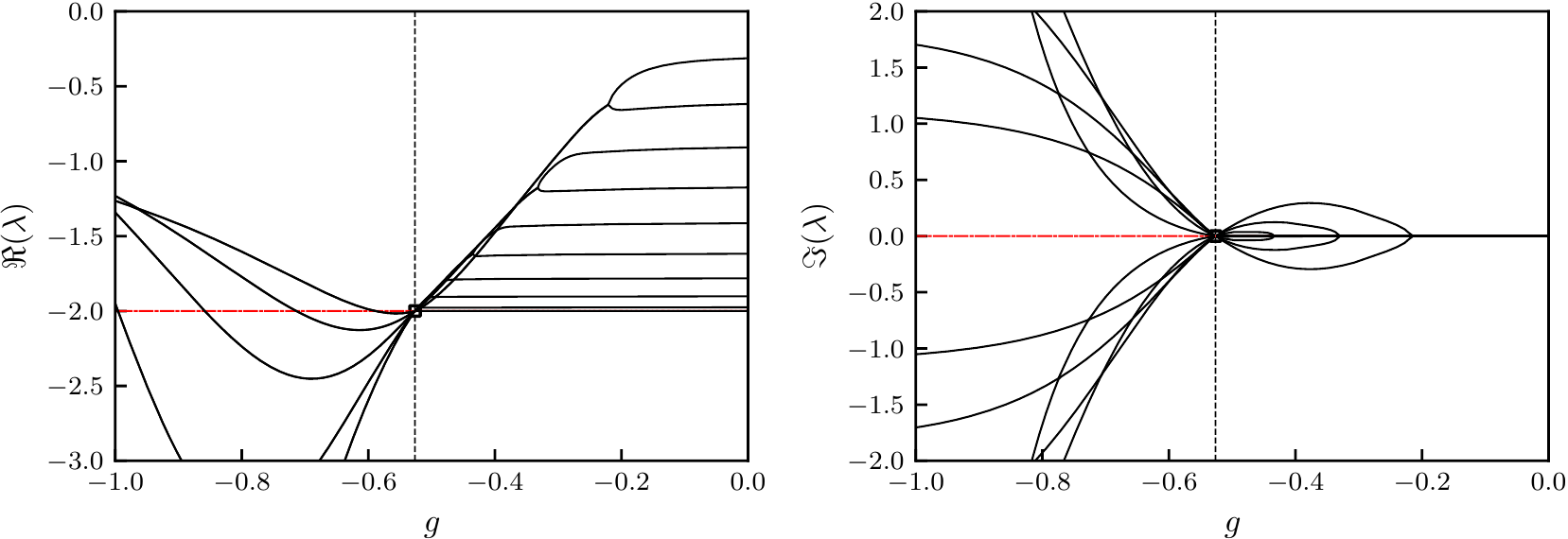}
\caption{Real and imaginary parts of the rapidities $\{\lambda_1 \dots
\lambda_N\}$ for a system with the parameters of Figure \ref{fig:RG:ener} at half-filling with $L=20$, $N=10$ and varying $g=GL$. The vertical line marks $g_c = G_cL$, while the horizontal red line marks $\lim_{k \to 0} \epsilon_k = -2(t_1+t_2)$. \label{fig:RG:rap}}
\end{center}
\end{figure}

The solutions to the Richardson-Gaudin equations for the ground state of the presented Hamiltonian are given in Fig. \ref{fig:RG:rap}. These always arise as either real variables or in complex pairs, and it can be seen that for $G = G_c$, all solutions collapse to a single point $\lambda_{\alpha} = -c_0/c_1=-2(t_1+t_2)$ corresponding to the non-interacting mode. Similar behaviour has been observed in $p$-wave \cite{ibanez_exactly_2009,rombouts_quantum_2010,links_exact_2015} and extended $d$-wave superconductors \cite{marquette_integrability_2013}. For $g>g_c$ a single variable $\lambda = -2(t_1+t_2) = \lim_{k \to 0} \epsilon_k $, unlike for $g<g_c$, indicating a change in the topological invariant $\lim_{k \to 0}n_k$ from $1$ to $0$.

\end{document}